\documentclass[a4paper,11pt]{article}
\usepackage{float}
\usepackage[pdftex]{graphicx}
\usepackage{subcaption}
\usepackage[T1]{fontenc}
\usepackage{lmodern}
\usepackage{slashed}
\usepackage[utf8]{inputenc}
\usepackage[english]{babel}
\usepackage{microtype}
\usepackage{cite}
\usepackage{amsmath,amssymb,amsfonts,amsthm}
\usepackage{mathtools,mathrsfs,calligra,aurical}
\usepackage[nottoc,notlot,notlof]{tocbibind}
\usepackage{upgreek}
\usepackage{mathtools}
\numberwithin{equation}{section}
\allowdisplaybreaks
\usepackage[all]{xy}
\usepackage{xcolor}
\usepackage{graphicx}
\graphicspath{{images/}}
\usepackage{geometry}
\usepackage[toc,page]{appendix}
\usepackage{hyperref}
\usepackage[normalem]{ulem}

\usepackage{bm}
\usepackage{ragged2e}
\usepackage{appendix}
\usepackage{slashed}
\usepackage{bbold}
\usepackage{cancel}

\definecolor{blue-violet}{rgb}{0.54, 0.17, 0.89}
\definecolor{PineGreen}{cmyk}{0.92, 0, 0.59, 0.25}
\definecolor{YellowOrange}{cmyk}{0, 0.42, 1, 0}
\definecolor{orange}{rgb}{0.95, 0.5, 0.1}
\newcommand{\dd}{\mathrm{d}}

\DeclareMathAlphabet{\mathpzc}{OT1}{pzc}{m}{it}

\usepackage[dvipsnames]{xcolor}

\begin{document}

\begin{titlepage}
\begin{flushright}
\par\end{flushright}
\vskip 0.5cm
\begin{center}
\textbf{\LARGE \bf  Multi-Boundary Spinning $AdS_5$ Black Holes, Strongly Coupled Plasma Balls and a dual locally de Sitter spacetime}\\
\vskip 5mm

\vskip 1cm

\large {\bf Andr\'{e}s Anabal\'{o}n,}$^{~a}$\footnote{anabalo@gmail.com} \large {\bf Guillermo Buckle,}$^{ ~a}$\footnote{gbuckle2022@udec.cl} \large {\bf Annelies Fullgraff,}$^{ ~a}$\footnote{afullgraff2022@udec.cl}  \\ \large {and \bf Marcelo Oyarzo}$^{ ~b}$\footnote{marceloandres.oyarzo@usc.es}

\vskip .5cm 

$^{(a)}${\textit{Departamento de F\'isica, Universidad de Concepci\'on, Casilla 160-C, Concepci\'on, Chile.}}\\ \vskip .1cm
$^{(b)}${\textit{Departamento de Física de Partículas $\&$ Instituto Galego de Física de Altas Enerxías (IGFAE),
Universidade de Santiago de Compostela, E-15782 Santiago de Compostela, Spain}}
\end{center}

\begin{abstract}
We construct a new family of exact spinning black hole solutions in five-dimensional General Relativity with a negative cosmological constant, which are dual to strongly coupled spinning plasma balls on four-dimensional Minkowski spacetime. In analogy with the recent discovery of such black solutions in four bulk dimensions, the boundary Lorentz factor acts as a new coordinate in the bulk, giving rise to a new UV region that, in our case, is locally a three-dimensional de Sitter spacetime times the real line. The dual energy-momentum tensor in this region describes a conformal anisotropic fluid that satisfies the weak and strong energy conditions.  
\end{abstract}
\vfill{}
\vspace{1.5cm}
\end{titlepage}

\setcounter{footnote}{0}
\tableofcontents

\section{Introduction and Discussion}
The discovery that the quark--gluon plasma (QGP) produced in non-central heavy-ion collisions can sustain extraordinarily large vorticities has opened up new avenues for investigating the properties of strongly interacting matter \cite{Becattini:2020ngo}. Measurements of the global polarization of $\Lambda$ hyperons at the Relativistic Heavy Ion Collider (RHIC) indicate that the QGP may constitute the most rapidly rotating fluid observed to date, with characteristic vorticities reaching $\omega\sim10^{21}\,\mathrm{s}^{-1}$ over fluid cells of typical subatomic dimensions, $R\sim10^{-15}\,\mathrm{m}$ \cite{STAR:2017ckg}. At the same time, the QGP exhibits another striking property: throughout the deconfined phase prior to hadronization, it is remarkably well described as a relativistic perfect fluid. This behavior is reflected in its exceptionally small shear-viscosity-to-entropy-density ratio, $\eta/s$, which is found to lie close to the universal lower bound obtained from holography \cite{Policastro:2001yc}.

Despite the substantial body of phenomenological evidence accumulated over the years, as well as the success of relativistic hydrodynamics in accounting for the large-scale evolution of the plasma, a fundamental theoretical description of strongly coupled rotating fluids is still incomplete. Furthermore, the situation is currently being revised for weakly coupled systems, for a recent discussion see \cite{Becattini:2025oyi}. 
In this paper we use holography\cite{Maldacena:1997re, Witten:1998qj} to shed light on this problem in four dimensions, extending the recent three-dimensional constructions of Refs. \cite{Anabalon:2026pha, Anabalon:2026iyj}. A rotating plasma at finite temperature is described holographically by means of an asymptotically AdS spinning black hole solution. The black hole must asymptotically match a conformal boundary with the natural representative of four dimensional Minkowski spacetime. We use coordinates
\begin{equation}
ds^2_r = -dt^2 + dw^2 + w^2 d\phi^2 + dz^2\, .
\end{equation}
At the same time, the holographic energy-momentum tensor of the dual field theory must have the form of a conformal perfect fluid in rigid rotation with constant angular velocity $\Omega_0$:
\begin{equation}
\langle T^r_{ij} \rangle = U_i U_j (\rho + P) + \chi^r_{ij} P\, ,
\end{equation}
where the four-velocity is given by $U = \gamma(\partial_t + \Omega_0 \partial_\phi)$, satisfying the normalization condition $U_i U^i = -1$, and the local Lorentz factor of the rigid rotation is defined by $\gamma = (1 - w^2 \Omega_0^2)^{-1/2}$.  Conformal invariance requires the energy-momentum tensor to be traceless, which fixes the equation of state of the plasma as $P = \rho / 3$. These conditions together with energy-momentum conservation completely fix the pressure to be proportional to $\gamma^4$. The constant of proportionality encodes the microscopic details of the system. For a detailed discussion, see Appendix \ref{app:perfect-fluid}.

We show that our background satisfies the requirements presented above, and we find that the pressure of the holographic fluid is
\begin{equation}
P = \frac{L^5 m \Omega_0^4}{2\kappa} \gamma^4 \,,
\end{equation}
where $m$ is an integration constant appearing in \ref{metric}, $L$ is the $AdS_5$ curvature radius, and $\kappa$ is the reduced Newton constant in five dimensions. Due to the physical constraint that the norm of the velocity cannot exceed the speed of light, the spinning plasma disk possesses a natural maximum spatial extent. Since the rotational speed depends on both the radial distance and the angular velocity, the speed of light is reached at the critical radius $w = \Omega_0^{-1}$, which corresponds to the divergence of the local Lorentz factor, $\gamma \to \infty$ . When fixing the conformal boundary at $r \to \infty$ to be flat Minkowski spacetime, we find that one of the non-compact bulk coordinates can be identified with the boundary Lorentz factor $\gamma$.
Furthermore, the limit $\gamma \to \infty$ defines a second asymptotic region of the five-dimensional bulk spacetime, in addition to the usual $r\to \infty$ region.
%Furthermore, we find that the region $\gamma = \infty$ extends naturally into the five-dimensional gravitational bulk as a new codimension-one conformal boundary located at the asymptotic region $y \to \infty$. 
This gives rise to a new conformal boundary that is locally three-dimensional de Sitter spacetime times the real line. We study the induced energy-momentum tensor in the de Sitter region and we find that it describes an anisotropic conformal fluid.

Hence, the asymptotically $AdS_5$ spacetime exhibits a remarkable two-boundary structure where two independent asymptotic regions coexist and join continuously. Their common boundary is reached as both non-compact holographic coordinates approach to infinity along any curve that keeps their ratio fixed, leading to an induces metric of the form
%of the form $r = C y$ with $y \to \infty$ and $C$ some constant. It is described by the metric:
\begin{equation}
 -\dd t^2 + \dd z^2 + \Omega_0^{-2} \dd\phi^2\, .
\end{equation}

Because the conformal boundary is four-dimensional, holographic renormalization involves a conformal anomaly \cite{Duff:1993wm}. We use the prescription of Ref. \cite{BirrellDavies1982} to isolate the traceless contribution to the energy-momentum tensor.

The paper is organized as follows. In Section 2, we present our conventions and the action principle. Section 3 introduces the new family of exact rotating solutions and presents a detailed analysis of their global geometric properties and the different limits. In Section 4, we examine the conformal boundaries of the spacetime, characterizing both the flat Minkowski boundary and the asymptotic $dS_3 \times \mathbb{R}$ geometry. In Section 5 the dual energy-momentum tensor is computed in the two asymptotic regions and their interpretation as the energy-momentum tensor of rigidly rotating plasma disk is analyzed. In Section 6 we provide some conclusions and future directions. 

\section{Setup}

We consider five-dimensional general relativity with a negative cosmological constant. The action principle is
\begin{equation}
S[g]=\frac{1}{2\kappa}\int d^5x \sqrt{-g}\left(R+\frac{12}{L^2}\right)+ \frac{1}{\kappa}\int d^4x\sqrt{-h}\left(K-\frac{3}{L} - \frac{L}{4} \mathsf{R}[h] \right) \,,
\end{equation}
where the last term is a boundary term, included to ensure that the variational principle is well defined and that the dual energy-momentum tensor is finite. Here $h$ is the determinant of the induced metric on the boundary, $K$ is the trace of the extrinsic curvature associated with the boundary hypersurface and $\mathsf{R}[h]$ its Ricci scalar\cite{Balasubramanian:1999re}. The field equations are%
\begin{align}
R_{\mu\nu}-\frac{1}{2}g_{\mu\nu}R -\frac
{6}{L^{2}}\,g_{\mu\nu}=0\,.
\end{align}

\section{The Geometry}
In this section we present a new metric given in the coordinate patch $(t,\phi,z,r,y)$. We extended the recent four-dimensional solution \cite{Anabalon:2026pha, Anabalon:2026iyj} following \cite{Chong:2004hw} 
\begin{align}\nonumber
ds^2& =
\alpha^2
\left(
\frac{
y^4 m
}{
y^2+r^2
}
-\frac{y^2 r^2}{L^2}
\right)\dd t^2
-
\frac{
2 \alpha y^2 (y^2-1)
m L^2
}{
y^2+r^2
}
\,\dd t\,\dd \phi\\\nonumber
&+
\left(
\frac{
(y^2-1)^2
m L^4
}{
y^2+r^2
}
+
L^2 (y^2-1)(r^2+1)
\right)\dd \phi^2\\ \label{metric}
&+
\frac{
L^2(y^2+r^2)
}{
y^2 (y^2-1)
}
\dd y^2
+
\frac{
y^2+r^2
}{\frac{r^4}{L^2}+\frac{r^2}{L^2}
-m
}
\dd r^2+\frac{\alpha^2}{L^2} r^2y^2 \dd z^2 \, .
\end{align}
In order to analyze the local regularity of the spacetime, let us consider the following vielbein basis
\begin{align}
e^{0} & =\frac{1}{Lr}\sqrt{\frac{\Delta}{\Sigma}}\boldsymbol{\Upsilon}\,, \qquad \label{vielbein}
e^{1}  =\frac{y}{L}\sqrt{\frac{y^{2}-1}{\Sigma}}\boldsymbol{\zeta}\,, \qquad 
e^{2}  =Lr\sqrt{\frac{\Sigma}{\Delta}}\dd r\,, \qquad \\
e^{3}  &=\frac{L\sqrt{\Sigma}}{y\sqrt{y^{2}-1}}\dd y\,,\qquad
e^{4}  =\frac{\alpha^{2}ry}{L^{2}}\dd z\,, \\
\boldsymbol{\Upsilon} & =\alpha y^{2}\dd t-L^{2}(y^{2}-1)\dd\phi\,, \qquad \boldsymbol{\zeta}  =-\alpha r^{2}\dd t+L^{2}(r^{2}+1)\dd\phi\,,
\end{align}
where $\Sigma = r^2 + y^2$ and $\Delta = r^6 + r^4 - mL^2 r^2$.

The components of the Riemann tensor in this orthonormal basis are
finite throughout the spacetime wherever $r^2+y^2>0$, to see its explicit expression see Appendix \ref{curvature-form}. In particular,
in the two asymptotic regions, the curvature two-form
\begin{equation}
\boldsymbol{R}^{ab}
=\frac{1}{2}R^{ab}{}_{cd}\,e^c\wedge e^d
\end{equation}
behaves as
\begin{align}
\lim_{r\to\infty}\boldsymbol{R}^{ab}
&=-\frac{1}{L^2}e^a\wedge e^b+\mathcal{O}(r^{-2}) \, ,\\
\lim_{y\to\infty}\boldsymbol{R}^{ab}
&=-\frac{1}{L^2}e^a\wedge e^b+\mathcal{O}(y^{-2}) \, .
\end{align}
Moreover, these limits are uniform along the asymptotic sectors
defined by
\begin{equation}
r=\lambda y \, ,
\end{equation}
for arbitrary fixed $\lambda$ and $y \to \infty$, for which
\begin{equation}
\boldsymbol{R}^{ab}
=-\frac{1}{L^2}e^a\wedge e^b+\mathcal{O}(y^{-2}) \, .
\end{equation}
Thus, the spacetime is asymptotically locally AdS in all asymptotic
directions, including the sectors in which the asymptotic region is
approached with $r/y$ held fixed. Nevertheless, the conformal
completion may fail to be differentiable at the locus where the
distinct asymptotic regions meet. We analyze this issue below.

The finiteness of the Riemann tensor in the orthonormal frame is also
reflected in the Kretschmann scalar,
\begin{equation}
R_{\mu\nu\rho\sigma}R^{\mu\nu\rho\sigma}
=
\frac{40}{L^4}
+
\frac{24m^2(3r^4-10r^2y^2+3y^4)}
{(r^2+y^2)^6}\, ,  
\end{equation}
which is finite for $r^2+y^2>0$. The determinant of the metric in this
coordinate patch is
\begin{equation}
\det g
=-\alpha^4L^2r^2y^2(r^2+y^2)^2 \, .
\end{equation}

The normalization of the metric has been chosen such that $\phi$ is a
canonically normalized angular coordinate, with
\begin{equation}
\phi\sim\phi+2\pi \, .
\end{equation}
The vector $\partial_\phi$ degenerates at $y=1$, which therefore
defines the natural lower endpoint of the $y$-coordinate. Since we are
primarily interested in black-hole solutions, we restrict to $r>0$.
Thus, the coordinate ranges are
\begin{equation}
t\in\mathbb{R},
\qquad
y\geq1,
\qquad
\phi\in[0,2\pi),
\qquad
r>0 \, .
\end{equation}
Within this coordinate domain, the Kretschmann invariant and all
components of the Riemann tensor in the orthonormal frame are finite.
With the above periodic identification of $\phi$, the geometry is
smooth at $y=1$.

The black-hole horizons are determined by the zeros of
\begin{equation}
F(r) \equiv  L^2(r^2+y^2)g^{rr}
=r^4+r^2-mL^2.
\end{equation}
The positive root is
%\begin{equation}
%r_+^2
%=-\frac{1}{2}
%+\frac{1}{2}\sqrt{1+4mL^2},
%\end{equation}
%or
\begin{equation}
r_+
=
\left[
-\frac{1}{2}
+\frac{1}{2}\sqrt{1+4mL^2}
\right]^{1/2}\, ,
\end{equation}
for $r>0$. Hence, for $m>0$, there is a single event horizon at $r=r_+$. For
$m=0$, the horizon degenerates to $r_+=0$.

Since $\phi$ is the only periodic coordinate, possible closed timelike
curves can arise only if the corresponding closed orbits become
timelike. However,
\begin{equation}
g_{\phi\phi}
=
\frac{L^2(y^2-1)
\left[
L^2m(y^2-1)+(1+r^2)(r^2+y^2)
\right]}
{r^2+y^2}
>0
\end{equation}
for $m>0$ and $y>1$. Therefore, the closed $\phi$-orbits are everywhere
spacelike, and there are no closed timelike curves associated with the
periodic identification. No closed timelike curves at large values of y requires that $mL^2>-1$.

The metric \eqref{metric} admits a closed conformal Killing--Yano tensor and therefore possesses a Carter-like constant for geodesic motion. For explicit expressions and an analysis of the geodesic equations in the asymptotic regions, see Appendix \ref{app:hidden-sym-geodesic}.

For $m=0$, the spacetime is locally AdS with radius $L$.
%, since
%\begin{equation}
%R^{\mu\nu}{}_{\rho\sigma}
%=
%-\frac{1}{L^2}
%\left(
%\delta^\mu_\rho\delta^\nu_\sigma
%-\delta^\mu_\sigma\delta^\nu_\rho
%\right).
%\end{equation}
Indeed, introducing
\begin{equation}
\rho=\alpha ry,
\qquad
w=
\frac{L\sqrt{(r^2+1)(y^2-1)}}{\alpha ry},
\qquad
z=ZL,
\end{equation}
brings the metric to the Poincar\'e-patch form
\begin{equation}
ds^2_{\mathrm{AdS}}
=
-\frac{\rho^2}{L^2}dt^2
+\frac{L^2}{\rho^2}d\rho^2
+\rho^2
\left(
dw^2+w^2d\phi^2+dZ^2
\right).
\end{equation}
In this limit, $r_+=0$ corresponds to the Poincar\'e horizon.

The generator of the horizon at $r=r_+$ is
\begin{equation}
\xi=\partial_t+\omega\partial_\phi \, ,
\qquad
\omega
=
\frac{\alpha r_+^2}
{L^2(1+r_+^2)} \,.
\end{equation}
where $\omega$ is the horizon angular velocity. The Hawking temperature is
\begin{equation}
T=\frac{\kappa_{\mathrm{grav}}}{2\pi} \, ,
\qquad
\kappa_{\mathrm{grav}}
=
\left.
\sqrt{
-\frac{1}{2}
\nabla^\mu\xi^\nu
\nabla_\mu\xi_\nu
}
\right|_{r=r_+} \, ,
\end{equation}
where $\kappa_{\mathrm{grav}}$ is the surface gravity at the horizon. For the present solution, the temperature is
\begin{equation}
T=
\frac{\alpha r_+(1+2r_+^2)}
{2\pi L^2(1+r_+^2)}\, .
\end{equation}

Finally, the horizon area density is
\begin{equation}
\mathcal{A}
=
\left.
\sqrt{g_{yy}g_{\phi\phi}g_{zz}}
\right|_{r=r_+}
=
\alpha L r_+(1+r_+^2)y \, .
\end{equation}
Since the horizon is non-compact, its total area is infinite; hence,
$\mathcal{A}$ should be interpreted as the area density with respect
to the non-compact directions.
 
\subsection{Static limit}
Let us consider the following suitable change of coordinates and reparameterization of $m$,
\begin{equation}
    r = \frac{\rho}{\alpha(1 + \frac{\alpha^2 \hat{x}^2}{2 L^2}) } \, , \qquad y = 1 + \frac{\alpha^2 \hat{x}^2}{2 L^2} \, , \qquad z = L  \hat{z} \, , \qquad m = \frac{m_s}{\alpha^4} \,.
\end{equation}
The slowly rotating limit $\alpha \to 0$ is
\begin{equation}
    \mathrm{ds^2} =-\left(\frac{\rho^2}{L^2}-\frac{m_s}{\rho^2}\right)\dd t^2 + \frac{ \dd \rho^2}{\frac{\rho^2}{L^2}-\frac{m_s}{\rho^2}}+\rho^2(dx^2+ x^2 \dd \phi^2 + \dd \hat{z}^2) -\frac{2m_sx^2\alpha}{\rho^2} \dd \phi \dd t\,.
\end{equation}
When $\alpha=0$ this is the standard planar Schwarzschild-AdS black hole.

\subsection{Region $r^2<0$}
The spacetime can be analytically extended beyond the region $r^2\geq 0$ into a region with $r^2<0$. In this extended region, the curvature becomes singular at
\begin{equation}
r^2+y^2=0 \, .
\end{equation}
This behavior is reminiscent of the analytic extension of the Myers--Perry geometry, where the radial coordinate can likewise be continued to a region with $r^2<0$, with curvature singularities 
appearing at specific loci in this extended region \cite{Gibbons:2009um}. Indeed, the leading form of the metric around $r=0$ is
\begin{align}
   \mathrm{ds^2} = L^2\left[(y^2-1)\dd \phi^2 + \frac{\dd y^2}{y^2-1}\right] + m \left[ \alpha y \dd t - \frac{L^2(y^2-1) \dd \phi}{ y} \right]^2 + y^2 \left(-\frac{\dd r^2}{m} + \frac{\alpha^2}{L^2} r^2 \dd z^2\right) \, .
\end{align}
The first bracket corresponds to the hyperbolic space $H_2$ with coordinates $(y,\phi)$. The first and second terms form a 3-dimensional space $\Sigma_3$ that can be thought of as a line bundle over $H_2$, where the fiber coordinate is $t$. In the parentheses of the last term, there is a Minkowski spacetime in Milne coordinates covering the interior of the past (future) light cone. The surface $r\to0$ corresponds to the light-cone surface. Therefore, close to $r=0$, the space is regular and acquires the form of a warped product of $\Sigma_3$ and 2-dimensional Minkowski space.

Now we will analytically continue the space to the region with $r^2<0$. We thus consider the change of coordinates $r=\sqrt{x}$ in the metric \eqref{metric}, leading to
\begin{align}\label{metric_r0}
   \mathrm{ds^2} = \frac{\alpha^2 x y^2 \dd z^2}{L^2}+\frac{L^2 \dd x^2 \left(x+y^2\right)}{4 x \left(-L^2 m+x^2+x\right)} + \dots
\end{align}
where the ellipses stand for the remaining terms of the metric \eqref{metric}. In these coordinates, $x$ can take negative values. 
The possible horizons are given by the roots of the polynomial
\begin{equation}
x \left(-L^2 m+x^2+x\right)=0 \label{thepolynomial}
\end{equation}
which are
\begin{align}
x_0=0 \, , \qquad x_{\pm} = \frac{1}{2}(-1 \pm \sqrt{1 + 4mL^2}) \, .
\end{align}
The ordering of the above values depends on the sign of $m$. Indeed, for $m>0$, we have
\begin{align}
x_-<-1 \,, \qquad x_+>0 \,, \quad \forall m>0 \,.
\end{align}
While for $m<0$, the spacetime is geodesically complete for $r\geq 0$ because the Killing vector $\partial_z$ vanishes at $r=0$. Regularity requires that $z$ be compact. Thus, these spaces always hide a singularity.

\section{Conformal Boundary}

The geometry possesses two independent asymptotic regions associated with the non-compact coordinates $r$ and $y$. The conformal boundary is constructed from the union of these regions. In this section, we characterize these regions purely from a geometric viewpoint before discussing their holographic interpretation.

On one hand, we find the standard Minkowski conformal boundary at $r\to \infty$. Defining the conformal factor
\begin{equation}
     W(r,y) = \frac{L}{\alpha r y} \label{W_factor}
\end{equation}
we find that
\begin{align}
    \dd s^2_r\equiv\lim_{r \to \infty} W(r,y)^{2} \dd s^2 =   -\dd t^2 + \dd w^2 + w^2 \dd \phi ^2 + \dd z^2 \,,
\end{align}
where we have used the new coordinate $w$ defined by
\begin{align}
    y = \frac{1}{\sqrt{1 - \frac{\alpha^2 w^2}{L^4}}} \, .
\end{align}
The other asymptotic region yields
\begin{align}\label{mety}
\dd s^2_y\equiv \lim_{y \to \infty} W(r,y)^{2} \dd s^2=&
-\left(1-\frac{mL^2}{r^2}\right)\dd t^2
-\frac{2mL^4}{\alpha r^2}\,\dd t\,\dd \phi
+\frac{L^4\,\dd r^2}
{\alpha^2\left(-mr^2L^2+r^6+r^4\right)}\nonumber \\
&+
\left[
\frac{L^4}{\alpha^2}
+\frac{L^4(mL^2+1)}{\alpha^2r^2}
\right]\dd \phi^2 
+\dd z^2 \, .
\end{align}
The spacetime \eqref{mety} is the direct product between  a space of constant curvature and the real line parameterized by $z$. Indeed, we find that the three-dimensional metric satisfies
\begin{align}
R^{t r}{}_{t r} &= \frac{\alpha^2 m}{L^2} \,,\\
R^{t\phi}{}_{t\phi} &= \frac{\alpha^2 m}{L^2}\, ,\\
R^{r\phi}{}_{r\phi} &= \frac{\alpha^2 m}{L^2}\, .
\end{align}
Therefore, black holes (solutions with $m>0$) have a conformal boundary that is locally de Sitter times the real line at $y=\infty$. These two regions of the conformal boundary meet at their common boundary. This can be found by setting $r=Cy$ in the bulk and then taking $y\to\infty$. The result is independent of $C$ and is given by

\begin{align}
    \lim_{y \to \infty} W(r=C y,y)^{2} \mathrm{ds}^2= -\dd t^2 + \dd z^2 +\Omega_0^{-2} \dd \phi^2 \,, \qquad \text{with} \quad \Omega_0=\alpha/L^2 \,.
\end{align}
$\Omega_0$ coincides with the angular velocity of a particle moving around the $w=0$ axis with speed $w\Omega_0$. Hence, $y(w)$ is the Lorentz dilation factor. In terms of the coordinate $y$ the Minkowski boundary is 
\begin{equation}
\mathrm{ds}_r^2=-\dd t^2+\Omega_0^{-2}\frac{\dd y^2}{(y^2-1)y^4}+\Omega_0^{-2}(y^2-1)\frac{\dd \phi^2}{y^2} + \dd z^2\, .
\end{equation}
Thus, the conformal boundary metric is continuous at $y=\infty=r$.

In the next section, we shall construct the dual energy-momentum tensor in the different conformal boundaries that we have discussed.

\section{Holography} \label{sec:holography}
We pick the representative of the conformal boundary as $\chi^I_{i j}$
\begin{align}
    \chi^I_{ij} \dd x^i \dd x^j = \mathrm{ds}_I^2 \, ,
\end{align}
for $I=r,y$. The renormalized energy-momentum tensor is given by
\begin{align}
    \langle T^I_{i j} \rangle &= -\frac{2}{\sqrt{-\chi^I}} \frac{\delta S_{\rm ren}}{\delta \chi^{I i j}} = -\lim_{I\to \infty} W(r,y)^{-2} \frac{2}{\sqrt{-h}} \frac{\delta S_{\rm ren}}{\delta h^{i j}}\\
    &=  \lim_{I \to \infty} \frac{W(r,y)^{-2}}{\kappa} \left( -K_{i j} + h_{i j} K -\frac{3}{L}h_{i j} + \frac{L}{2} \mathsf{G}_{i j} \right) \, , \label{Tij_holographic}
\end{align}
where the indices $i,j$ run over the tangent space of the boundary space, namely $h_{ij}$ is the non-degenerate part of $h_{\mu \nu} = g_{\mu \nu} - n_\mu n_\nu$ and $n_\mu$ is the spacelike unit normalized vector that is orthogonal to the boundary. The extrinsic curvature is defined as $K_{\mu \nu} = h_{\mu}{}^{\rho} h_{\nu}{}^{\sigma} \nabla_{\rho} n_{\sigma}$ and $K$ as its trace. The tensor $\mathsf{G}_{i j}$ is Einstein's tensor computed with the metric $h_{i j}$.

Our analysis requires the use of several standard results from four dimensional QFT that we reproduce here for the ease of reading the article.
In four dimensions, the general expression of the conformal anomaly 
is \cite{Duff:1993wm, BirrellDavies1982}
\begin{equation}
\langle T_{i}^{i}\rangle =\frac{c}{16\pi ^{2}}C^{2}-\frac{a}{%
16\pi ^{2}}E_4\text{ ,}
\end{equation}%
where $C^{2}$ is the square of the Weyl tensor and $E_4$ is the Euler density.
For free fields the QFT calculation yields a result in terms of the number
of scalars $N_{S}$, number of Dirac fermions $N_{F}$ and number of gauge
vectors $N_{V}$: 
\begin{align}
c& =\frac{1}{120}\left( N_{S}+6N_{F}+12N_{V}\right) \text{ ,} \\
a& =\frac{1}{360}\left( N_{S}+11N_{F}+62N_{V}\right) \text{ .}
\end{align}%
When the bulk Lagrangian is Einstein gravity with a comological constant the
result is \cite{Henningson:1998gx}:

\begin{equation}
c=a=\frac{N^{2}}{4}\text{ .}
\end{equation}%
In the large $N$ limit, there is full agreement with the $SU(N)$, $\mathcal{N%
}=4$ Super Yang-Mills field content, which is given by $N_{S}=6\left(
N^{2}-1\right) $, $N_{F}=2\left( N^{2}-1\right) $ and $N_{V}=\left(
N^{2}-1\right) $. Indeed, $\mathcal{N}=4$ Super Yang-Mills is not a free
theory, but its Weyl anomalies exactly match the result in terms of free
fields. In terms of gravity variables the central charges are
\begin{equation}
c=a=\frac{\pi^2 L^{3}}{\kappa}\text{ .}
\end{equation}%

We shall use the following result that yields the transformation rule between conformally related metrics $g_{i j}$ and $\bar{g}_{i j}$ with $a=c$\footnote{This can be deduced from the discussion of section 6 of \cite{ BirrellDavies1982}.},
\begin{equation}\label{anomalytrans}
\langle T^i_{j} [\boldsymbol{g}]\rangle = \left( \frac{\bar{g}}{g}\right)^{1/2}\left( \langle T^i_{j}[\boldsymbol{\bar{g}}] \rangle + \frac{a}{16 \pi^2} H^i_{j}[\boldsymbol{\bar{g}}]\right)-\frac{a}{16 \pi^2}H^i_{j}[\boldsymbol{g}] 
\end{equation}
with
\begin{align}
    H_{ij} = 2(R_i{}^k R_{kj} -\frac23 R_{ij}R -\frac12 R_{kl}R^{kl} g_{ij} + \frac14 R^2 g_{ij}) \, .
\end{align}
This equation is valid when $\bar{g}$ is a conformally flat metric. In this case is possible to check that
\begin{align}
H_{i}^i=E_4=
R_{ijkl}R^{ijkl}
-4R_{ij}R^{ij}
+R^2\, . 
\end{align}
Let us analyze the dual energy-momentum tensor at the two asymptotic regions.

\subsection{Dual energy-momentum tensor at $r= \infty$}
To carry on the holographic renormalization program, we first compute the dual energy-momentum tensor with a conformal factor
\begin{equation}
\frac{L}{\alpha r} \, ,
\end{equation}
and then we make a Weyl transformation to pick the representative with
\begin{equation}
W(r,y)=\frac{L}{\alpha r y}\, .
\end{equation}

The first step yields an energy-momentum tensor in a background that is conformally Minkowski spacetime
\begin{equation}
y^2 \dd s_r^2=y^2\chi^r_{i j}\dd x^i \dd x^j=\bar{\chi}^r_{i j}\dd x^i \dd x^j  \, .
\end{equation}
for which the boundary metric is
\begin{align}
    \bar{\chi}^r_{ij}\dd x^idx^j = \frac{1}{1-\alpha^2 w^2/L^4} (-\dd t^2+\dd w^2 +w^2 \dd \phi ^2 + \dd z^2) \, . \label{metric_gamma_r_inf}
\end{align}
The dual energy-momentum tensor is given by
\begin{align}
    \langle T^r_{ij}[\bar{\chi}^r] \rangle = T^{\mathrm{anomaly}}_{ij} + T^{(m)}_{ij} \label{full_Tij_r_infinity}
\end{align}
it is divergence-free with respect to the metric $\bar{\chi^r}_{ij}$ and piece $T^{(m)}_{ij}$ is traceless. Their explicit expression are
\begin{align}
    T_{ij}^{\mathrm{anomaly}}\dd x^{i}\dd x^{j} & =\frac{\alpha^{4}\left(-4L^{8}-4\alpha^{2}L^{4}w^{2}+\alpha^{4}w^{4}\right)}{8\kappa L\left(L^{4}-\alpha^{2}w^{2}\right)^{3}}(-\dd t^{2}+\dd z^{2}) +  \notag \\
    & \quad +\frac{(\alpha^{8}w^{4}-4\alpha^{6}L^{4}w^{2})\dd w^{2}-(4\alpha^{6}L^{4}w^{4}+3\alpha^{8}w^{6})\dd \phi^{2}}{8\kappa L\left(L^{4}-\alpha^{2}w^{2}\right)^{3}} \\
T_{ij}^{(m)}\dd x^{i}\dd x^{j} & =\frac{\alpha^{4}Lm}{2\kappa\left(L^{4}-\alpha^{2}w^{2}\right)^{2}}\left\{ \dd t^{2}\left(3L^{4}+\alpha^{2}w^{2}\right)+w^{2}\dd \phi^{2}\left(L^{4}+3\alpha^{2}w^{2}\right)\right.\\
 & \quad\left.+\left(\dd w^{2}+\dd z^{2}\right)\left(L^{4}-\alpha^{2}w^{2}\right)-8\alpha L^{2}w^{2}\dd t \dd \phi\right\} 
\end{align}
We use now \eqref{anomalytrans} to find the energy-momentum tensor in the Minkowski background
\begin{align}
    \langle T^r_{ij}[\chi^r] \rangle = \gamma^2 T^{m}_{ij}\equiv T^{\rm fluid}_{ij}
\end{align}
where $T^{\rm fluid}$ is the energy-momentum tensor of a perfect fluid defined in flat spacetime with metric $\chi^r_{i j}$
\begin{align}
    T^{\rm fluid}_{ij} = \mathsf{U}_i \mathsf{U}_j (\rho + P) + \chi^r_{i j} P
\end{align}
with
\begin{align}
    P = \frac{\rho}{3} = \frac{ L^5 m \Omega_0 ^4}{2 \kappa }\gamma^4
  \,, \qquad
    \mathsf{U} = \gamma (\partial_t + \Omega_0 \partial_ \phi) , \qquad
    \mathsf{U}^i \mathsf{U}_i = -1 \,, \qquad
    \gamma = \frac{1}{\sqrt{1 - \Omega_0^2 w^2}} \,. 
\end{align}
Hence, it is traceless and conserved in Minkowski spacetime.
\subsection{Dual energy-momentum tensor at $y= \infty$}
Again, we first construct the energy-momentum tensor with the conformal factor
\begin{align}
    \frac{L}{\alpha y}
\end{align}

\begin{align}
    \bar{\chi^y}_{ij}\dd x^i \dd x^j = \lim_{y\to \infty} U(y)^2 g_{ij}\dd x^i \dd x^j 
     &= - \frac{r^4+r^2 -m L^2}{1+L^2 m + r^2}\dd t^2 + \frac{L^4 \dd r^2}{\alpha^2 (r^4 + r^2 -m L^2)}  + r^2 \dd z^2 + \\
    & \quad + \frac{L^4 (1+L^2 m +r^2)}{\alpha^2}\left( \dd \phi - \frac{\alpha m}{1+L^2 m +r^2} \dd t \right)^2\, . \notag
\end{align}
This is just a conformally rescaled three-dimensional de Sitter spacetime times the real line \eqref{mety}. Along the same lines as in the previous section, we find the energy-momentum tensor in the background \eqref{mety}, namely
\begin{align}
\langle T^y_{ij}[\chi^y] \rangle = r^2 T^{m}_{ij}\, ,
\end{align}
which yields
\begin{align}
    \langle T^y_{ij}[\chi^y] \rangle = \frac{ r^2  }{2\kappa}\footnotesize
\left(
\begin{array}{cccc}
-\dfrac{\alpha^{4}m\left(L^{2}m-r^{2}-4r^{4}\right)}{L^{3}}
&
\dfrac{\alpha^{3}m\left(L^{2}m-4r^{2}-4r^{4}\right)}{L}
&
0
&
0
\\[1.2em]
\dfrac{\alpha^{3}m\left(L^{2}m-4r^{2}-4r^{4}\right)}{L}
&
-\alpha^{2}Lm\left(-3+L^{2}m-7r^{2}-4r^{4}\right)
&
0
&
0
\\[1.2em]
0
&
0
&
-\dfrac{\alpha^{4}mr^{2}}{L^{3}}
&
0
\\[1.2em]
0
&
0
&
0
&
\dfrac{\alpha^{2}Lm}{L^{2}m-r^{2}-r^{4}}
\end{array}
\right) \normalsize
\end{align}
with coordinates $(t, \phi, z, r)$. It is possible to check that it is traceless and conserved in the background \eqref{mety}.

This is an anisotropic traceless fluid. We find the following decomposition in terms of eigenvectors of the energy-momentum tensor $\langle T^{yj}_{i}\rangle X^i=\lambda X^j$. The spacelikes eigenvectors are

\begin{align}
    X&=\partial_z\implies \lambda=P_z=-\frac{m\alpha^4 r^4}{2\kappa L^3} \, ,\\
    X&=\partial_r\implies \lambda=P_r=-\frac{m\alpha^4 r^4}{2\kappa L^3} \, ,\\
X&=r(\partial_t+\Omega_0\partial_{\phi})\implies \lambda=P_{\phi}=\frac{3 m\alpha^4 r^4}{2\kappa L^3} \,.
\end{align}
The timelike vector yields
\begin{equation}
V=r^3\frac{\alpha^2}{L^4}\left(\frac{\alpha L^2 (1+r^2)}{r^2}\partial_t+\partial_{\phi}\right)\implies -\lambda=\rho=\frac{ m\alpha^4 r^4}{2\kappa L^3}
\end{equation}
Hence, it satisfies the weak and strong energy conditions.

\section{Conclusions}
In this paper we constructed a new class of solutions to the five-dimensional Einstein equations with a negative cosmological constant. These solutions provide a holographic description of strongly coupled spinning plasma disks in four-dimensional Minkowski spacetime. They have no CTCs and there are no curvature singularities outside the horizon. The geometry has two asymptotic AdS regions $r\to \infty$ and $y\to \infty$, corresponding to the UV of the RG flow and the ultrarelativistic limit, respectively. An upshot of the description is that typically the ultrarelativistic limit of a spinning plasma is singular. However, by means of holography, we constructed a regular description of it in terms of an anisotropic traceless fluid that satisfies the weak and strong energy conditions and lives in three-dimensional de Sitter space times the real line. We find remarkable that when the black D3 branes spins in the brane (and not in the $S^5$), naturally creates a curved D3 brane with a de Sitter region. 

We showed that in the static limit we recover the black D3 brane. When the spacetime is rotating, we found that is possible to extend the spacetime into the region of $r^2 <0$ where there is a singularity. The configurations with negative masses are regular provided $z$ is compactified. However, these spaces must be studied to clarify whether they suffer of instabilities due to nucleation of bubbles of nothing \cite{Witten:1981gj} or are more like a ground state along the AdS soliton like behaviour \cite{Horowitz:1998ha}. We found that the requirement of no closed timelike curves bounded the mass parameter $mL^2>-1$. 

Models of plasma balls can be found in \cite{Nastase:2005rp, Emparan:2009dj, Aharony:2005bm}, although they do not include spin. Importantly, however, they capture the fact that the RHIC plasma ball ends before reaching the speed of light. In our model, this is achieved by introducing a boundary (rather than a conformal boundary) at a finite value of $y$. A holographic description of CFTs with boundaries was proposed in \cite{Takayanagi:2011zk}, which we believe provides a natural framework to pursue in order to obtain a more realistic description of a plasma ball, starting from our model.

The supersymmetric limits of the configurations presented here imply $m=0$, which is just $AdS_5$ in oblate coordinates. In a future work, we intend to extend the present analysis to the charged case. This will allow us to study non-trivial possible supersymmetric solutions to further understand the structure of these configurations and their connection with the physics of the strongly coupled spinning plasmas.

\section*{Acknowledgements}
The work of AA is supported in part by the FONDECYT grants 1230853, 1242043, 1250133, 1262452 and 1262414.  MO is supported by AEI-Spain PID2023-152148NB-I00 and by Maria de Maeztu excellence unit grant CEX2023-001318-M, by Xunta de Galicia (CIGUS Network of Research Centres and project ED431F-2023/19), and by the European Union FEDER. The work of GB is supported by the National Agency for Research and Development (ANID) through the National Master's Scolarship 22262144
\appendix 
\section{Perfect fluid kinematics with spin}
\label{app:perfect-fluid}

In this appendix, we provide some details on how the non-anomalous contribution to the holographic energy-momentum tensor obtained in Sec.\ref{sec:holography} can be interpreted as the energy-momentum tensor of a rigidly rotating perfect fluid. In particular, we show that the conservation of the energy-momentum tensor, together with conformal invariance, completely determines its dependence on the coordinates up to an overall constant.

Consider Minkowski spacetime in cylindrical coordinates,
\begin{equation}
    \dd \bar{s}^{2}
    =
    -\dd t^{2}+\dd w^{2}+w^{2}\dd \phi'^{2}+\dd z^{2} \,.
\end{equation}
To describe a rigidly rotating fluid with constant angular velocity $\Omega$, it is convenient to introduce the co-rotating coordinate
\begin{equation}
    \phi'=\phi+\Omega t \,,
\end{equation}
so that the metric becomes
\begin{equation}
    \dd \bar{s}^{2}
    =
    -\left(1-\Omega^{2}w^{2}\right)\dd t^{2}
    +2\Omega w^{2}\dd t\,\dd \phi
    +w^{2}\dd \phi^{2}
    +\dd w^{2}+\dd z^{2}.
    \label{eq:rotating-flat-metric}
\end{equation}
We introduce the Lorentz factor
\begin{equation}
    \gamma=\frac{1}{\sqrt{1-\Omega^{2}w^{2}}} \,.
    \label{eq:gamma-app}
\end{equation}
In these coordinates the fluid is at rest, and its four-velocity is
\begin{equation}
    u=\gamma\,\partial_t,
    \qquad
    u^\mu u_\mu=-1 \,.
    \label{eq:fluid-velocity-app}
\end{equation}

To identify the physical quantities measured in the local rest frame of the fluid, we introduce an orthonormal frame. In the plane spanned by $\partial_t$ and $\partial_\phi$, we look for a spacelike vector $v$ orthogonal to $u$ and normalized according to
\begin{equation}
    v^\mu v_\mu=1,
    \qquad
    u^\mu v_\mu=0 \,.
\end{equation}
These conditions determine
\begin{equation}
    v
    =
    \gamma w\Omega\,\partial_t
    +\frac{1}{\gamma w}\,\partial_\phi \, .
\end{equation}
Hence, the natural orthonormal vielbein can be chosen as
\begin{align}
    e^{\hat{0}}
    &=
    -\sqrt{1-\Omega^{2}w^{2}}\,\dd t
    +\frac{\Omega w^{2}}{\sqrt{1-\Omega^{2}w^{2}}}\,\dd \phi \,,
    \\
    e^{\hat{1}}
    &=\dd w \,,
    \\
    e^{\hat{2}}
    &=
    \frac{w}{\sqrt{1-\Omega^{2}w^{2}}}\,\dd \phi \,,
    \\
    e^{\hat{3}}
    &=\dd z \,,
\end{align}
which satisfies
\begin{equation}
    d\bar{s}^{2}
    =
    \eta_{\hat a\hat b}\,
    e^{\hat a}e^{\hat b},
    \qquad
    \eta_{\hat a\hat b}
    =
    \mathrm{diag}(-1,1,1,1) \,.
\end{equation}

The advantage of this frame is that the hydrodynamic interpretation of the energy-momentum tensor becomes manifest. Assuming that the state is isotropic in the local rest frame of the fluid, the energy-momentum tensor must be diagonal in the orthonormal frame,
\begin{equation}
    T_{\hat a\hat b}
    =
    \mathrm{diag}
    \left(
        \rho(w),P(w),P(w),P(w)
    \right).
    \label{eq:T-orthonormal}
\end{equation}
Equivalently, in the coordinate basis it takes the perfect-fluid form
\begin{equation}
    T_{\mu\nu}
    =
    \left(\rho+P\right)u_\mu u_\nu
    +P\,\bar{\gamma}_{\mu\nu} \,.
    \label{eq:perfect-fluid-app}
\end{equation}
The components in the orthonormal and coordinate bases are related through the vielbein according to
\begin{equation}
    T_{\mu\nu}
    =
    e^{\hat a}{}_{\mu}
    e^{\hat b}{}_{\nu}
    T_{\hat a\hat b} \,.
    \label{eq:T-vielbein-transformation}
\end{equation}

Since the boundary theory is conformal, the non-anomalous part of the energy-momentum tensor must be traceless,
\begin{equation}
    T^\mu{}_\mu=0 \,.
\end{equation}
In four spacetime dimensions this implies
\begin{equation}
    -\rho+3P=0 \,,
\end{equation}
and therefore
\begin{equation}
    P=\frac{\rho}{3} \,.
    \label{eq:conformal-eos-app}
\end{equation}
Thus, after imposing isotropy and conformal invariance, the only remaining function to be determined is $\rho(w)$.

This function is fixed by conservation of the energy-momentum tensor,
\begin{equation}
    \bar{\nabla}_{\mu}T^{\mu\nu}=0 \,.
    \label{eq:conservation-app}
\end{equation}
For the Ansatz above, all components of the conservation equation are identically satisfied except for the radial one. The latter reduces to
\begin{equation}
    \left(1-\Omega^{2}w^{2}\right)
    \frac{\dd \rho}{\dd w}
    -
    4\Omega^{2}w\,\rho=0 \,.
    \label{eq:rho-ode}
\end{equation}
Its solution is
\begin{equation}
    \rho(w)
    =
    \frac{\rho_{0}}
    {\left(1-\Omega^{2}w^{2}\right)^{2}}
    =
    \rho_{0}\gamma^{4} \,,
    \label{eq:rho-gamma4}
\end{equation}
where $\rho_{0}$ is an integration constant. Consequently,
\begin{equation}
    P(w)
    =
    \frac{\rho_{0}}{3}\gamma^{4} \,.
\end{equation}

Therefore, the $\gamma^{4}$ dependence appearing in the holographic energy-momentum tensor is not an accidental feature of the gravitational solution. Rather, it is precisely the radial dependence required by conservation for a conformal perfect fluid undergoing rigid rotation in four spacetime dimensions. Comparison with the holographic energy-momentum tensor fixes the integration constant to be
\begin{equation}
    \rho_{0}
    =
    \frac{3L^{5}m\Omega^{4}}{2\kappa},
\end{equation}
and hence
\begin{equation}
    P=\frac{\rho}{3},
    \qquad
    \rho
    =
    \frac{3L^{5}m\Omega^{4}}{2\kappa}\,
    \gamma^{4} \,,
\end{equation}
all in agreement with the result presented in the main text.

\section{Curvature 2-form} \label{curvature-form}

The components of the curvature 2-forms are expressed in a convenient way in terms of the functions
\begin{align}
    \Sigma = r^2 + y^2 \,, \quad X=\frac{r^2-y^2}{\Sigma} \,, \quad Y = \frac{2 r y}{\Sigma} \,, \quad \mu = \frac{m L^2}{\Sigma^2} \,,
\end{align}
\begin{align}
    L^2\boldsymbol{R}^{01} &= -e^0\wedge e^1 - \mu (X e^0 \wedge e^1-Y e^2 \wedge e^3) \, ,
    \\
    L^2\boldsymbol{R}^{23} &= -e^2\wedge e^3 - \mu (X e^2 \wedge e^3 + Y e^0 \wedge e^1) \, ,
    \\
    L^2\boldsymbol{R}^{03} &= -e^0\wedge e^3 + \mu (-X e^0 \wedge e^3 + Y e^1 \wedge e^2) \, ,
    \\
    L^2\boldsymbol{R}^{12} &= -e^1\wedge e^2 - \mu (X e^1 \wedge e^2 + Y e^0 \wedge e^3) \, ,
    \\
    L^2\boldsymbol{R}^{02} &= -e^0\wedge e^2 + \mu [(1+2X) e^0 \wedge e^2 + 2 Y e^1 \wedge e^3] \,, 
    \\
    L^2\boldsymbol{R}^{13} &= -e^1\wedge e^3 + \mu [(-1+2X) e^1 \wedge e^3 - 2 Y e^0 \wedge e^2] \,,
    \\
    L^2\boldsymbol{R}^{a4} &= -(1+(-1)^a \mu )e^a\wedge e^4 \, , \quad a=0,1,2,3 \,.
\end{align}
in the vielbein basis defined in \eqref{vielbein}. All components of the curvature 2-form are regular for $r^2+y^2>0$.

\section{Hidden symmetry and geodesic separability} \label{app:hidden-sym-geodesic}

The spacetime with metric \eqref{metric} admits three independent Killing vectors,
\begin{align}
\partial_t\,,\qquad \partial_\phi \,,\qquad \partial_z \,.
\end{align}
In addition to these manifest isometries, the geometry possesses a non-trivial hidden symmetry encoded in a closed conformal Killing--Yano (CKY) 2-form $\boldsymbol{Y}$. In five dimensions, a closed CKY 2-form $\boldsymbol{Y} = \frac12 Y_{\mu \nu} \dd x^\mu \dd x^\nu$ satisfies
\begin{align}
\nabla_\rho Y_{\mu\nu} = \frac12 g_{\rho[\mu|}\nabla_\sigma Y^\sigma{}_{|\nu]} \,,
\qquad
\dd\boldsymbol{Y}=0 \,,
\end{align}
see, e.g., \cite{Frolov:2006dqt}. Since $\boldsymbol{Y}$ is closed, it can locally be written as
\begin{align}
\boldsymbol{Y}=\dd\boldsymbol{P} \,.
\end{align}
A solution of the CKY equation is given by the following potential
\begin{align}
\boldsymbol{P} = \frac{L^{2}}{2\alpha} \left(1-y^{2}-r^{2}y^{2}\right)\dd\phi +\frac{r^{2}y^{2}}{2}\dd t \,,
\end{align}
which leads to
\begin{align}
	\boldsymbol{Y} = & \frac{L^{2}r(y^{2}-1)}{\alpha}\dd\phi\wedge\dd r
	+\frac{L^{2}(1+r^{2})y}{\alpha}\dd\phi\wedge\dd y
	+ry^{2}\dd r\wedge\dd t
	-r^{2}y\dd t\wedge\dd y \, .
\end{align}

The Hodge dual of a closed CKY tensor is a Killing-Yano tensor. In the present case this gives the Killing-Yano 3-form
\begin{align}
\boldsymbol{f}\equiv\star\boldsymbol{Y}
={}&
ry\left[
-Ly(1-y^{2})\dd\phi\wedge\dd r
-Lr(1+r^{2})\dd\phi\wedge\dd y 
+\frac{\alpha y^{3}}{L}\dd r\wedge\dd t
+\frac{\alpha r^{3}}{L}\dd t\wedge\dd y
\right]\wedge\dd z \,.
\end{align}
Its square defines a rank-two Killing tensor. For later convenience, we choose the equivalent representative
\begin{align}
K^{\mu\nu} = \frac{\alpha^{2}}{2} f^{\mu\rho\sigma}f^{\nu}{}_{\rho\sigma}
	+\frac{L^{2}}{\alpha^{2}}\ell^\mu\ell^\nu \,,
\end{align}
where the second term is a reducible Killing tensor constructed from
\begin{align}
\boldsymbol{\ell} &\equiv \partial_t+\Omega_0\partial_\phi
= -\frac{1}{\alpha^{2}} \nabla^\rho Y_{\rho}{}^\mu\partial_\mu \,,
\qquad 
\Omega_0=\frac{\alpha}{L^{2}} \,.
\end{align}
The vector $\boldsymbol{\ell}$ is conformally related to the vector of the perfect fluid in both asymptotic regions. At this end of this section, we will analyze some properties of this vectors field. Explicitly, the Killing tensor is
\begin{align}
K^{\mu\nu} ={}&
    \frac{1}{y^{2}-1}\delta_\phi^\mu\delta_\phi^\nu
    -\frac{L^{4}}{\alpha^{2}y^{2}}\delta_t^\mu\delta_t^\nu
    -L^{2}y^{2}g^{\mu\nu}
    +y^{2}(y^{2}-1)\delta_y^\mu\delta_y^\nu \, .
\end{align}

This non-trivial Killing tensor provides a Carter-like constant and allows us to separate the Hamilton-Jacobi equation for geodesic motion. We shall use this Carter constant to analyze the null geodesics that could reach asymptotic regions $r\to \infty$ and $r\to \infty$. Writing
\begin{align}
g^{\mu\nu}p_\mu p_\nu=-\epsilon \,,
\qquad
p_\mu=\partial_\mu S \,,
\end{align}
and using the Ansatz
\begin{align}
S=-et + j\phi + kz + S_r(r) + S_y(y) \,,
\end{align}
the additional conserved quantity is
\begin{align}
\mathcal{C}=K^{\mu\nu}p_\mu p_\nu \,.
\end{align}
The Hamilton-Jacobi equation then separates into
\begin{align}
(p_{y})^{2}+\frac{1}{(y^{2}-1)}\left[L^{2}\epsilon-\frac{\mathcal{C}}{y^{2}}+\frac{j^{2}}{y^{2}(y^{2}-1)}-\frac{e^{2}}{\Omega_{0}^{2}y^{4}}\right]&=0 \,,\\
(p_{r})^{2}+\frac{r^{2}\mathcal{C}+L^{2}r^{4}\epsilon}{\Delta(r)}-\frac{r^{4}}{\Omega_{0}^{2}\Delta(r)^{2}}[(e-\Omega_{0}j)^{2}mL^{2}+r^{2}(e^{2}-\Omega_{0}^{2}j^{2})+e^{2}]&=0\,.
\end{align}
We recall the definition $\Delta(r)=r^6 + r^4 -m L^2 r^2$. For null geodesics, $\epsilon=0$, the asymptotic form of these equations is
\begin{align}
    (p_y)^2
    -\frac{\mathcal C}{y^4}
    +\frac{-\mathcal C+j^2-e^2/\Omega_0^2}{y^6}
    +\mathcal O(y^{-8})
    &=0\,,
\\
    (p_r)^2
    +\frac{\mathcal C}{r^4}
    +\frac{-\mathcal C+j^2-e^2/\Omega_0^2}{r^6}
    +\mathcal O(r^{-8})
    &=0\,.
\end{align}
The sign of the Carter constant therefore distinguishes the two asymptotic regions: geodesics reaching $y\to\infty$ require $\mathcal C\geq0$, whereas those reaching $r\to\infty$ require $\mathcal C\leq0$. In the critical case $\mathcal C=0$, the leading obstruction vanishes in both directions and accessibility is controlled by the subleading terms. In particular, at this order both limits require
\begin{align}
    \frac{e^2}{\Omega_0^2}-j^2\geq0\,.
\end{align}

An interesting locus in the asymptotic region is defined by the vanishing norm of the Killing vector $\boldsymbol{\ell}$ that, as we mentioned, is related by a conformal factor to the vector fields of the perfect fluids living in the different asymptotic region. Its norm is given by
\begin{align}
    \ell^\mu\ell^\nu g_{\mu\nu} &= \frac{L^2\Omega^2}{r^2+y^2} \Upsilon(r,y)\,,
\\
    \Upsilon(r,y) &\equiv L^2m-r^2-r^4+y^2(y^2-1)\,.
\end{align}
Indeed,
\begin{align}
    \ell^\mu\ell^\nu g_{\mu\nu} <0\,, \quad r\rightarrow\infty\,,\\
    \ell^\mu\ell^\nu g_{\mu\nu} >0\,, \quad  y\rightarrow\infty\,,
\end{align}
so that $\Upsilon=0$ separates regions in which $\boldsymbol{\ell}$ is timelike and spacelike. We therefore refer to this locus as the speed-of-light surface associated with $\boldsymbol{\ell}$. Solving $\Upsilon=0$ gives
\begin{align}
    y^2
    =
    \frac12\left[
    1+\sqrt{(1+2r^2)^2-4L^2m}
    \right]\,.
\end{align}
At large $r$ this curve approaches the corner $r,y\rightarrow\infty$ with $ \frac{y}{r}\to 1$.

Notice that the speed-of-light surface is not itself null. Its normal has norm
\begin{align}
(\dd\Upsilon)^2
=
\frac{
(2r^3+r)^2(-L^2m+r^4+r^2)
+(y^2-1)(1-2y^2)^2y^4
}{
L^2(r^2+y^2)
}\,,
\end{align}
which is positive in the exterior region. Hence, the speed-of-light surface is a timelike hypersurface. This hypersurface could be interpreted as the interface between the two boundaries.

\hypersetup{linkcolor=blue}
\phantomsection % use it for correct TOC link !!!
\addtocontents{toc}{\protect\addvspace{4.5pt}}% add vertical space in TOC
\addcontentsline{toc}{section}{References} % add References to TOC
\bibliographystyle{mybibstyle}
\bibliography{bibliografia.bib}

\end{document}